\documentclass[preprint,showpacs,preprintnumbers,amsmath,amssymb]{revtex4}%
\usepackage{graphicx}
\usepackage{dcolumn}
\usepackage{bm}
\usepackage{amsmath}
\usepackage{epstopdf}
\usepackage{amsfonts}
\usepackage{amssymb}%
\providecommand{\U}[1]{\protect\rule{.1in}{.1in}}
\newcommand{\be}{\begin{equation}}
\newcommand{\en}{\end{equation}}
\newcommand{\bea}{\begin{eqnarray}}
\newcommand{\ena}{\end{eqnarray}}
\begin{document}
\title{Strange stars in $f(R)$ theories of gravity in the Palatini formalism}
\author{Grigoris Panotopoulos}
\email{grigorios.panotopoulos@tecnico.ulisboa.pt}
\affiliation{CENTRA, Instituto Superior T{\'e}cnico, Universidade de Lisboa,
Av. Rovisco Pa{\'i}s 1, Lisboa, Portugal}
\date{\today}

\begin{abstract}
In the present work we study strange stars in $f(R)$ theories of gravity in the Palatini
formalism. We consider two concrete well-known cases, namely the $R+R^2/(6 M^2)$ model as well as the $R-\mu^4/R$ model
for two different values of the mass parameter $M$ or $\mu$. We integrate the modified Tolman-Oppenheimer-Volkoff equations numerically, and we show
the mass-radius diagram for each model separately. The standard case corresponding to the General Relativity is also shown in the same figure for comparison.
Our numerical results show that the interior solution can be vastly different depending on the model and/or the value of the parameter of each model.
In addition, our findings imply that i) for the cosmologically interesting values of the mass scales $M,\mu$ the effect of modified gravity on strange
stars is negligible, while ii) for the values predicting an observable effect, the modified gravity models discussed here would be ruled out by their
cosmological effects.
\end{abstract}

\maketitle

\section{Introduction}

Current astrophysical and cosmological observational data indicate that the Universe is expanding at an accelerating rate.
The concordance model based on cold dark matter and a cosmological constant is the most economical model in excellent agreement
with the data. However, the problems with the cosmological constant have given birth to many alternatives, based either on a new dynamical field or on modifications to Einstein's General Relativity (GR), see e.g. \cite{copeland}. Of special interest are the models of $f(R)$ theories of gravity \cite{FofR}, which are a natural generalization of General Relativity and have been
extensively studied as alternatives to the cosmological constant to describe the current cosmic acceleration.

Compact objects \cite{textbook}, such as white dwarfs and neutron stars, are the final fate of stars.
The degeneracy pressure provided by the Fermi gas
balances the gravitational force, and the star finds a stable configuration. In white dwarfs the Fermi gas consists of
electrons, while in neutron stars the required pressure is provided by neutrons. Recently a new class of compact objects has
been postulated to exist due to some observed super-luminous supernovae \cite{superluminous}, which occur in about one out of
every 1000 supernovae explosions, and are more than 100 times brighter than normal supernovae. One plausible explanation is that neutrons are
further compressed so that a new object made of de-confined quarks is formed. This new compact object is called a "strange star", and since it is a much more stable configuration compared to a neutron star it could explain the origin of the huge amount of energy released in super-luminous
supernovae \cite{strangestars}. Compact objects, due to their unique properties, comprise an excellent natural laboratory to study and perhaps
constrain different models of modifications of gravity.

In the present work we wish to study strange stars in $f(R)$ theories of gravity. It is well known \cite{FofR} that in studying $f(R)$ theories of gravity
there are two different formalisms, namely the metric formalism and the Palatini formalism. In General Relativity they
are completely equivalent, however in modified theories of gravity they lead to different predictions. It is known that $f(R)$ theories of gravity
have difficulties with passing Solar System tests \cite{SST}, and also with supporting interior star
solutions \cite{stars}. In our work we have chosen to work in the Palatini formalism, since in this formalism the models are,
on the one hand, easily
compatible with Solar System tests, and on the other hand the interior star solution can match the exterior solution \cite{reijonen1, reijonen2}.
See, however, \cite{contraPalatini} for evidence against the cosmological and astrophysical viability of the
Palatini-$f(R)$ gravity.

We organize our work as follows: After this introduction, we present the theoretical framework in the next section.
In the third section we discuss two specific models and we present in figures our numerical results. Finally we conclude in the
last section.

\section{Strange stars in modified theories of gravity}

\subsection{Relativistic stars in General Relativity}

In this subsection we review relativistic stars, and in particular strange stars, in GR. The starting point is Einstein's
field equations with a cosmological constant
\be
G_{\mu \nu} + \Lambda g_{\mu \nu} = 8 \pi T_{\mu \nu}
\en
where we have set Newton's constant equal to unity, $G=1$, and in the exterior problem the matter energy-momentum
tensor vanishes. For matter we assume a perfect fluid with pressure $p$, energy density $\rho$ and an equation of state
$p(\rho)$. The energy-momentum trace is given by $T=-\rho+3p$. For the case of strange stars we shall consider the simplest equation of state corresponding to a relativistic gas of de-confined quarks, known also as the MIT bag model \cite{bagmodel}
\be
p = \frac{1}{3} (\rho - 4B)
\en
and the bag constant has been taken to be $B=57 MeV/fm^{-3}$ (which is slightly smaller than the $59 MeV/fm^3$ required so that neutrons
do not coagulate into droplets of ud matter in the simple MIT bag model, see the book by Haensel, Potekhin, Yakovlev, 2007) \cite{book}).
We mention in passing that for refinements
of the bag model the reader could consult e.g. \cite{refine}, and for the present state-of-the-art the recent paper \cite{art}.
For the metric in the case of static
spherically symmetric spacetimes we consider the following ansatz
\be
ds^2 = -f(r) dt^2 + g(r) dr^2 + r^2 d \Omega^2
\en
with two unknown functions of the radial distance $f(r), g(r)$. For the exterior problem one obtains the well-known solution
\be
f(r) = g(r)^{-1} = 1-\frac{2 M}{r}-\frac{\Lambda r^2}{3}
\en
where $M$ is the mass of the star.
For the interior solution we introduce the function $m(r)$ instead of the function $g(r)$ defined as follows
\be
g(r)^{-1} = 1-\frac{2 m(r)}{r}-\frac{\Lambda r^2}{3}
\en
so that upon matching the two solutions at the surface of the star we obtain $m(R)=M$, where $R$ is the radius of the star.
The famous Tolman-Oppenheimer-Volkoff equations \cite{TOV} for the interior solution of a relativistic star with a
non-vanishing cosmological constant read \cite{prototype}
\bea
m'(r) & = & 4 \pi r^2 \rho(r) \\
p'(r) & = & - (p+\rho) \: \frac{m(r)+4 \pi p r^3-\frac{\Lambda r^3}{3}}{r^2 (1-\frac{2 m(r)}{r}-\frac{\Lambda r^2}{3})}
\ena
where the prime denotes differentiation with respect to r, and the equations are to be integrated with the initial conditions $m(r=0)=0$ and $p(r=0)=p_c$, where $p_c$ is
the central pressure. The radius of the star is determined requiring that the pressure vanishes at the surface,
$p(R) = 0$, and the mass of the star is then given by $M=m(R)$. Strange stars with the above equations of state in GR have been
investigated in \cite{prototype} for negative and positive values of the cosmological constant. The observed value of the cosmological
constant is too small to have an effect. Only when $\Lambda$ is comparable to the bag constant B has an observable effect.

\subsection{Relativistic stars in $f(R)$ in the Palatini formalism}

We shall consider models described by an action of the form
\be
S = \int d^4x \sqrt{-g} \left (\frac{f(R)}{8 \pi} + \mathcal{L}_M \right )
\en
where $f(R)$ is a generic function of the Ricci scalar, and $\mathcal{L}_M$ is the matter Lagrangian
density. In the Palatini formalism the field equations take the form \cite{reijonen1}
\be
F(R) R_{\mu \nu} - \frac{1}{2} f(R) g_{\mu \nu} = 8 \pi T_{\mu \nu}
\en
where $F(R)=df(R)/dR$ is the first derivative of $f(R)$ with respect to $R$. Note that when
$f(R)=R$ and $F(R)=1$ the field equations are reduced to the standard equations of GR.
Taking the trace of the above field equations we obtain an algebraic equation relating the Ricci scalar
$R$ to the trace of the matter energy-momentum tensor $T=-\rho+3p$ as follows \cite{reijonen1}
\be
R F(R)-2f(R) = 8 \pi T
\en
In cases where the trace T does not depend on r, the field equations take the familiar form corresponding
to Einstein's equations with a non-vanishing cosmological constant, namely \cite{reijonen1}
\be
G_{\mu \nu} + \Lambda_{\rho} g_{\mu \nu} = \frac{8 \pi T_{\mu \nu}}{F_{\rho}}
\en
where the effective cosmological constant is given by
\be
\Lambda_{\rho} = \frac{1}{2} \left (R-\frac{f(R)}{F(R)} \right)
\en
which is evaluated at the value $R_0$ that solves the trace equation. The same holds for $F_{\rho}$.
For a generic equation of state where the trace T is a function of the radial distance r, one must integrate
the full (more complicated) field equations as was done in \cite{reijonen2} for polytropic stars. If we
set $f(r)=e^{A(r)}$ and $g(r)=e^{B(r)}$, then the differential equations
for the functions $A(r),B(r)$ are the following \cite{reijonen1, reijonen2}:
\begin{eqnarray}
A' & = & -\frac{1}{1+\gamma} \left( \frac{1-e^B}{r}+\frac{\alpha}{r}-8 \pi r p \frac{e^B}{F} \right) \\
B' & = & \frac{1}{1+\gamma} \left( \frac{1-e^B}{r}+\frac{\alpha+\beta}{r}+8 \pi r \rho \frac{e^B}{F} \right)
\end{eqnarray}
where $\gamma=(rF')/(2F)$, while the $\alpha, \beta$ are given by
\begin{eqnarray}
\alpha & = & r^2 \left(\frac{3}{4} \left(\frac{F'}{F} \right)^2+\frac{2 F'}{r F}+\frac{e^B}{2} \left(R-\frac{f}{F} \right) \right) \\
\beta & = & r^2 \left(\frac{F''}{F}-\frac{3}{2} \left(\frac{F'}{F} \right)^2  \right)
\end{eqnarray}
Returning to the equation of state $p=(1/3) (\rho-4B)$, for the exterior problem the energy-momentum vanishes,
while the effective cosmological constant becomes $\Lambda_0=\Lambda_{\rho}(T=0)$. Therefore, in the Palatini
formalism for a constant energy-momentum trace the problem is reduced to the usual problem of GR with a non-vanishing
cosmological constant, and with new pressure and energy density for the matter fluid $\tilde{p}=p/F_{\rho}$ and $\tilde{\rho}=\rho/F_{\rho}$.
The modified pressure and energy density satisfy the same equation of state, but now the bag constant is modified by the same
factor, namely $\tilde{B}=B/F_{\rho}$. Therefore we have to integrate the TOV equations valid in GR with the substitution
$B \rightarrow \tilde{B}$. Finally, one more difference compared to GR comes from matching the two solutions (interior and exterior problems),
since in $f(R)$ theories of gravity there are two different cosmological constants
outside the star and inside the star. This implies that the mass of the star is now given by \cite{reijonen2}
\be
M=m(R)+\frac{(\Lambda_{\rho}-\Lambda_0) R^3}{6}
\en

\section{Numerical results in two concrete models}

Here we shall consider two specific models, namely the Starobinsky model \cite{starobinsky}, $f(R)=R+R^2/(6 M^2)$
as well as the $1/R$ model that has been used to describe the current cosmic acceleration \cite{trodden}, $R-\mu^4/R$.
Each model is characterized by a single parameter which is a mass scale. In Starobinsky's model the corrections to GR
are important in the early universe, and we can have a successful inflationary model if $M \sim 10^{12} GeV$, while in the
second model the corrections are important at late times, and the mass scale $\mu \sim H_0 \sim 10^{-33} eV$ \cite{trodden}, where $H_0$ is 
today's Hubble parameter. In the cosmologically interesting range of the parameters $M,\mu$ the effect on the strange star is negligible. In 
this work, however, the mass scales are just two free parameters and we shall not consider the aforementioned values. Strange stars 
in $R^2$ gravity have also been studied in \cite{kokkotas, strangename}, but in the Einstein frame, while in this article we work in 
the Jordan frame and in the Palatini formalism.

a) Starobinsky model: For a model of the form $f(R)=R+R^2/(6 M^2)$ the functions we need are found to be
\bea
R_0 & = & 32 \pi B \\
F_{\rho} & = & 1+\frac{32 \pi B}{3 M^2} \\
\Lambda_{\rho} & = &  \frac{256 \pi^2 B^2}{32 \pi B + 3 M^2}  \\
\Lambda_0 & = & 0
\ena
since for strange stars the energy-momentum trace is $T=-4 B$. We have considered two different values of $M$
for which $F_{\rho}=3/2$ and $F_{\rho}=2$. In Fig.~1 below we show the mass-radius diagram for Starobinsky's
model for $M=\sqrt{\frac{64 \pi B}{3}}$ and for $M=\sqrt{\frac{32 \pi B}{3}}$, and for comparison we also show
the standard results corresponding to GR.

b) $1/R$ model: In this case the relevant functions are computed to be
\bea
R_0 & = & 16 \pi B + \sqrt{3 \mu^4 + 256 \pi^2 B^2}\\
F_{\rho} & = & 1+\frac{\mu^4}{(16 \pi B+\sqrt{256 \pi^2 B^2+3 \mu^4})^2} \\
\Lambda_{\rho} & = &  \frac{\mu^4 (16 \pi B+\sqrt{256 \pi^2 B^2+3 \mu^4})}{\mu^4+(16 \pi B+\sqrt{256 \pi^2 B^2+3 \mu^4})^2}  \\
\Lambda_0 & = & \frac{\sqrt{3} \mu^2}{4}
\ena
We have considered two different values for the mass scale $\mu$ for which $F_{\rho}=1.05511$ and $F_{\rho}=1.11112$.
In Fig.~2 below we show the mass-radius diagram for the two values $\mu = 3 \sqrt{\pi B}$ and $\mu = 4 \sqrt{\pi B}$
together with the GR results for comparison. The numerical results shown in Figures 1 and 2 clearly show that the
interior solution of strange stars can be vastly different depending on the model and/or the value of the parameter of each model.

\begin{figure}[!htb]
\centering
\includegraphics[scale=.9]{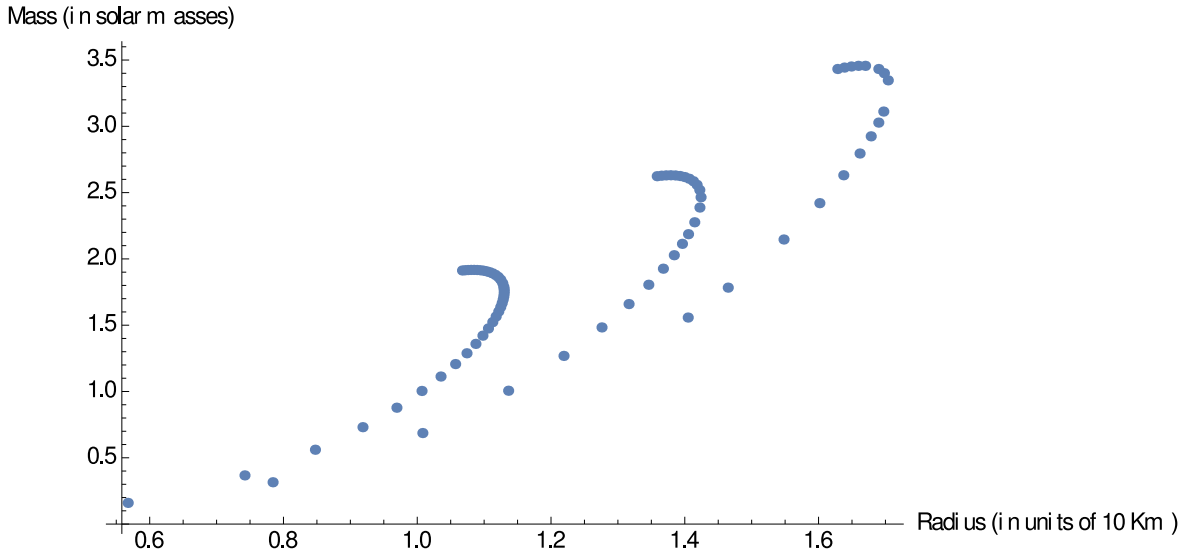}
\caption{Mass-radius diagram for quark stars in GR and in the Starobinsky model for two
different values of the mass scale M. From left to right $F=1$ (GR), $F_\rho=3/2$, $F_\rho=2$.}
\end{figure}

\begin{figure}[!htb]
\centering
\includegraphics[scale=1]{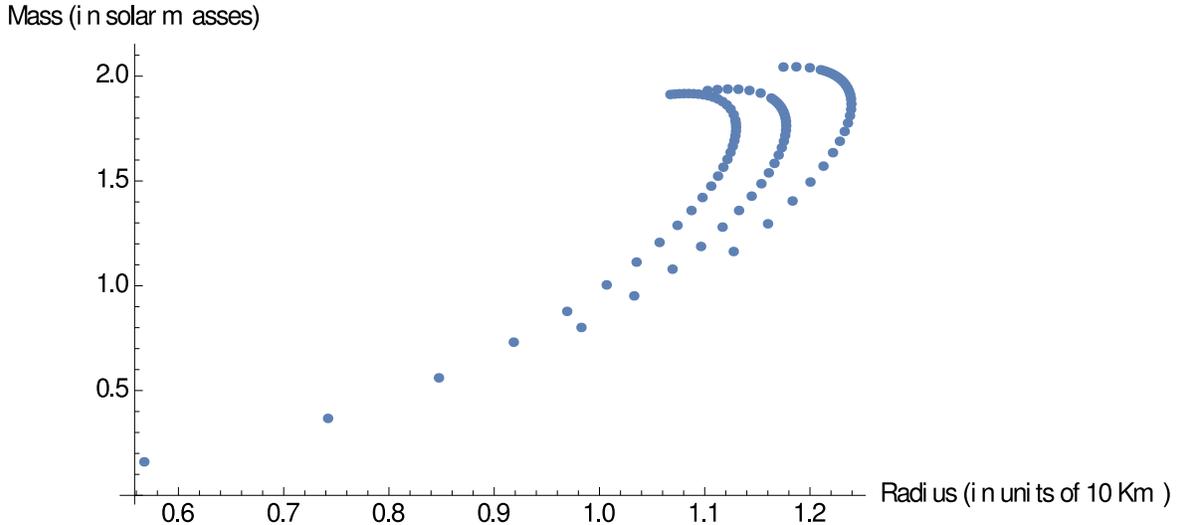}
\caption{Same as in Fig.~1 but for the model $R-\mu^4/R$. From left to right $F=1$ (GR), $F_\rho=1.05511$, $F_\rho=1.11112$.}
\end{figure}

A final comment is in order before finishing. Following \cite{prototype,kokkotas,strangename} we have considered here the simplest equation of 
state corresponding to the bag model, namely radiation plus a bag constant. If a refined equation of state is used \cite{refine} the trace of 
the energy-momentum tensor, as stated before, will not be a constant any more, and the full field equations must be integrated. In this case 
one expects that the predictions of the same $f(R)$ models will change. It would be interested to see exactly how different the new predictions 
will be compared to the results obtained here, and we wish to be able to address that issue in a future work.

\section{Conclusions}\label{conclu}

To summarize, in the present work we have studied strange stars in $f(R)$ theories of gravity in the Palatini
formalism. On the one hand it is known that this class of modified theories of gravity can explain the current
cosmic acceleration without a cosmological constant and also without introducing a new dynamical field.
On the other hand strange stars are hypothetical compact objects, like neutron stars, which are made out
of de-confined quarks. This composition in theory leads to a more stable configuration, and thus could explain the huge amount
of energy released in the observed super-luminous supernovae, which are more than 100 times brighter than normal supernovae.
We have considered the simple equation of state $p=(1/3) (\rho-4B)$ from the MIT bag model, which in the framework of
Einstein's General Relativity has already been investigated. Here we have considered two concrete well-known $f(R)$ models, namely
the $R+R^2/(6 M^2)$ model as well as the $R-\mu^4/R$ for two different values of the mass parameter $M$ or $\mu$.
We have integrated the modified Tolman-Oppenheimer-Volkoff equations numerically for many different values of the central pressure
as an initial condition, and finally we have shown the mass-radius diagram for each case separately for two different values of the mass 
parameter $M$ or $\mu$. For comparison the standard diagram corresponding to General Relativity is also shown in the same figure. Our 
numerical results show that the interior solution can be vastly different depending on the model and/or the value of the parameter 
of each model. In addition, our findings imply that i) for the cosmologically interesting values of the mass scales $M,\mu$ the effect of 
modified gravity on strange stars is negligible, while ii) for the values predicting an observable effect, the modified gravity models 
discussed here would be ruled out by their cosmological effects.


\begin{acknowledgments}
The author wishes to thank the reviewers for valuable comments and suggestions. His work was support 
from "Funda{\c c}{\~a}o para a Ci{\^e}ncia e Tecnologia".
\end{acknowledgments}


\end{document}